# *GTI-mSEMP Framework : A Proposed Framework to Simulate Malware Propagation with Inclusion of Attacker-Defender Strategy*

**Shadeeb Hossain[1,2,†] [ORCID ID: 0000-0002-5224-7684],**

**Elias Calboreanu[2] [0009-0008-9194-0589],**

**Kristopher Wilson[2]**

*[1] Shadeeb Engineering Lab, Research Division, Brooklyn, NY 11223.*

*[2]Capitol Technology University, Department of Engineering, Laurel, MD 20708*

***Abstract :***

Automated, multi-vector malware campaigns pose severe operational risks to heterogeneous, resource-constrained cyber-physical and wireless sensor networks. Conventional epidemiological models typically rely on time-invariant transition rates and flat topologies, failing to capture strategic interactions between adaptive adversaries and budget-constrained defenders across distinct architectural layers. To address this limitation, this paper proposes the Game-Theory-Integrated Modified Multi-Wireless Sensor Epidemic Malware Propagation (GTI-mSEMP) framework across a three-tier hierarchy comprising Gateway, Routing, and Peripheral nodes under Brute-Force Attack (BFA) and Zero-Day vectors. By coupling compartmental Susceptible-Exposed-Infectious-Recovered (SEIR) dynamics with a bi-level Stackelberg game, offensive and defensive resource allocations dynamically modulate transition probabilities, payload execution latency, recovery rates, re-susceptibility, and the basic reproduction number ($R_0$). Analytical stability and equilibrium properties are established using the Next Generation Matrix method, Routh-Hurwitz criteria, and existence/uniqueness results over compact strategy spaces. Numerical simulations using 4th-order Runge-Kutta integration and Sequential Quadratic Programming demonstrate that dynamic Stackelberg control outperforms static and perimeter-only postures under balanced scaling ($\eta \geq \mu$), whereas introducing exponential strategy smoothing ($\alpha = 0.3$) under an attacker-dominant regime ($\mu > \eta$) mitigates solver chattering, shifts defense toward high-density Routing and Peripheral tiers, and prevents catastrophic network collapse.

[†] Corresponding author: shadeeb@shadeebengineeringlab.com

## *I. Introduction*

With the rise of Internet of Things (IoT) connected smart cities, most wireless systems are now susceptible to malware attacks [1-4]. Some of the malware attacks include: (i) *worms, (ii) botnets, (iii) viruses, (iv)trojans, and (v)ransomware,* among others [5-8]. Malware can allow unauthorized access to the system and thereby contribute to several cybersecurity threats including leaking confidential information, consuming valuable resources or energy and even hostile takeover of the targeted network [9].

The malware spreading through the network can be modeled to assess their rate of compromise and predict their future stake in real time [10]. Cybersecurity analysts require dynamic simulated models to analyze the evolution and spread of malware code through their network.

In 2016, the Mirai botnet included a spree of large-scale denial of service (DDoS) attack [11]. Antonakakis et al. investigated the Mirai's botnets growth and DDoS activity between August 1, 2016, to February 28,2017 and created a comprehensive dataset to develop future technical and non-technical defenses for future attacks [12]. Another prominent botnet is the Mozi IoT malware attack in 2019 that was able to infect 1.5 million IoT devices. [13]. Unlike Mirai, which relies on a centralized command-and control (C2) server infrastructure, Mozi utilized a peer-to-peer (P2P) network topology based on a Distributed Hash Table (DHT).

Another example of a dangerous cyber threat is the zero-day ransomware Wannacry, that was able to halt the operations of large corporations and hospital facilities from over 150 countries [14]. Wannacry malware relies on the Server Message Block (SMB) protocol which checks the IP address and tries to connect over TCP port 445 [15-17].

Despite the security protocols, it becomes critical to have a comprehensive understanding of the propagation model in real-time to limit the spread. Mathematical simulations can play a critical role in such analysis. Susceptible, Exposed, Infected and Recovered (SEIR) are usually the four stages of a malware attack and the model that is used in predicting the malware propagation (particularly worm) in infected networks [18,10].

It is important to understand that most of the traditional compartmental models assume time-invariant transition rates as it only focuses on a particular attack-and-defense strategy. However, in real case scenarios, the defenders operate under strict operational and computational resource constraints, or the defender can actively change their approach depending on attackers' action. Stochastic games, which is a combination of game theory and Markov Decision Processes (MDP), can be used to analyze the randomness in the cyber-attack and therefore be included in the conventional SEIR model. Markov Decision Processes extend to a multiplayer competitive environment and can be used to characterize the randomness in cyber-attack and defense strategies [19]. This implies that the recovery rate ($\gamma$), transmission rate ($\beta$) and the effective reproduction number ($R_t$) cannot be treated as time-invariant constraints; instead, they function as dynamic variables dependent on the localized strategy profiles executed within the stochastic game framework.

The primary objective of this paper is to propose a Game-Theory-Integrated Modified Multi-Wireless Sensor Epidemic Malware Propagation (GTI-mSEMP) framework. This model accommodates a (i) *time-variant (attacker-defender response) recovery rate, (ii) time dependent (attacker-defender response) transmission rate,* and a (iii) *dynamic effective reproduction number*, thereby establishing a highly realistic simulation environment for active cyber defense-attack scenarios. The paper is divided into the following sections: **Section II**: *Relevant studies and background information*, **Section III**: *Proposed GTI-mSEMP framework*, **Section IV**: *Analytical Derivation of the Basic Reproduction Number,* **Section V**:*Proof of Equilibrium States and Game Theoretic Analysis,* **Section VI**: *Case Studies for Multi-Tier Architectural Hierarchy,* **Section VII:** *Simulation Setup, Numerical Results and Discussion*, **Section VIII**: *Computational Complexity, Limitations and Real-time Feasibility*, **Section IX**: *Conclusion*.

## *II. Relevant Studies and Background Information*

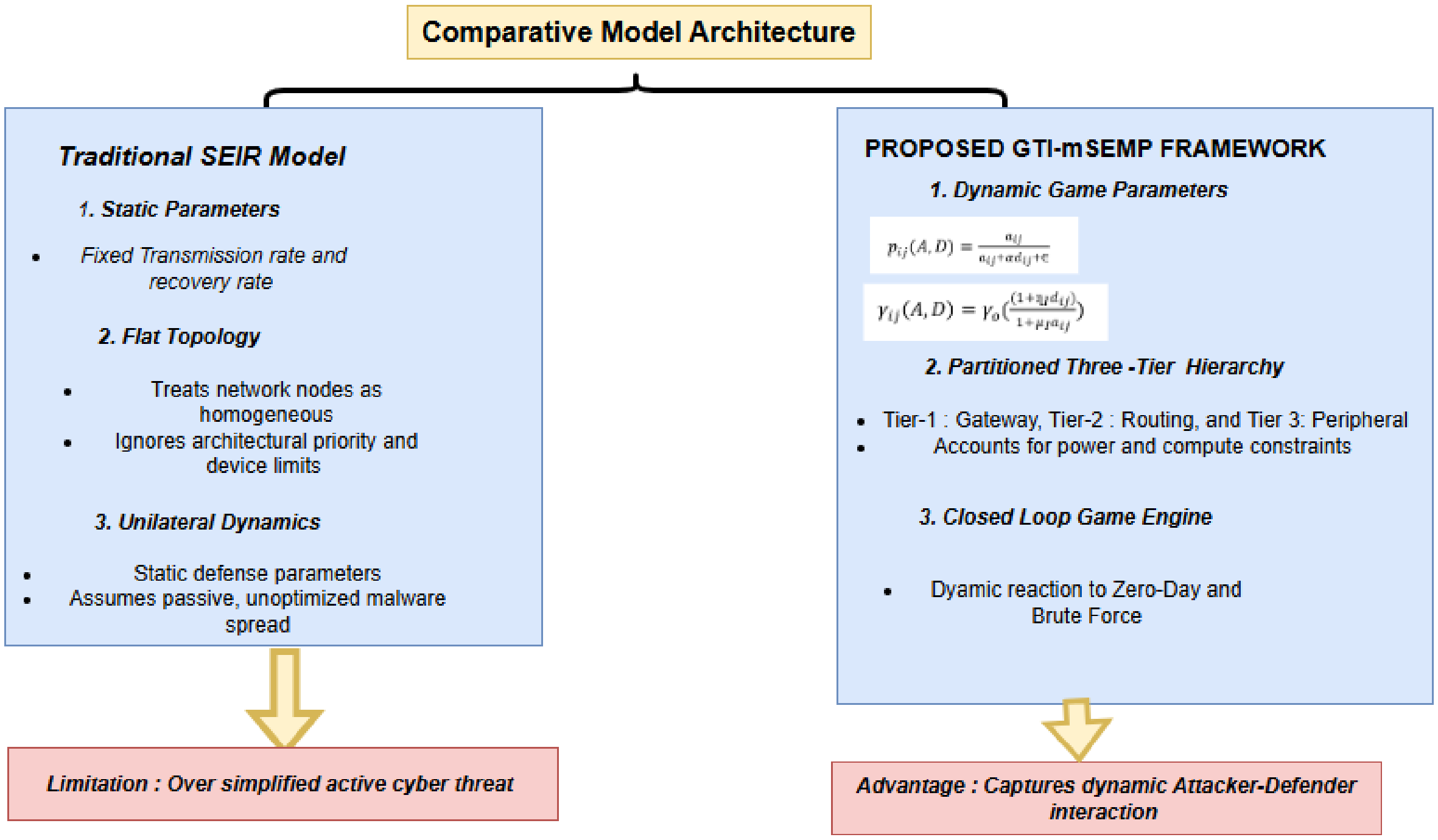

**Fig.1: Architectural comparison contrasting standard SEIR epidemic models against the proposed GTI-mSEMP framework.**

Epidemiological models have been widely adopted to characterize malware dissemination across wireless and IoT networks by mapping node states onto disease compartments such as Susceptible (S), Exposed (E), Infectious (I) and Recovered (R). Early and recent formulations -spanning fractional order virus models, non-linear quarantine compartments, bifurcation and stability analysis for WSN malware, and SEIRS and SEIRV variants with relapse or intervention provide foundational insight into outbreak thresholds and steady state containment [20-23].

The common state of the malware attack and defense mechanisms include:

(i) *Susceptible (S)* state: During this state, the wireless sensor nodes are still not exposed to the malware attack but are still susceptible to potential attack. The probability of susceptible nodes transitioning to exposed state is $p$. A stronger cybersecurity network will have a lower magnitude of $p$, compared to its counterparts.

(ii) *Exposed (E)* state : During this state, the sensor nodes are exposed to the malware and will be infected as a function of time unless an effective defense mechanism is adopted by the cybersecurity team of defense. Awasthi *et al.* (2023) proposed Exposed State-1($E_1$) and Exposed State -2 ($E_2$) to replicate two types of malware attack [20].

(iii) *Infected (I)* state: During this state, the infected sensor nodes can infect other neighboring sensor nodes. The rate of recovery, $\gamma$ , depend on both the defense and attacker's strategy and there is a possibility of the infected node to either transition to Recovered state or crash due to malfunction.

(iv) *Recovered (R)* state: The Recovered (R) state includes when the sensory nodes have actively recovered from their infectious state. The rate of re-susceptibility is given by $\delta$. This state is usually considered the last state of the cyber-attack cycle.

Fig.1 shows the architectural comparison between standard Susceptible-Exposed-Infected-Recovered (SEIR) against the proposed Game-Theory-Integrated Modified Multi-Wireless Sensor Epidemic Malware Propagation (GTI-mSEMP) framework. Classical compartmental SEIR formulations generally treat state transitions rates such as infection transmission, payload execution latency, and remediation patching as static, time-invariant hardware or software parameters [20-23]. Furthermore, standard models typically evaluate malware spread over flat, homogeneous topologies, ignoring the distinct computational and power constraints inherent to different physical device classes. In contrast the proposed GTI-mSEMP framework addresses these limitations by mapping dynamic state transitions directly to the real-time competitive cyber-attack scenario between adversary groups (attackers) and defenders. By coupling dynamic strategy allocation matrices with a strictly partitioned three-tier architectural hierarchy encompassing the framework captures realistic, non-linear epidemic trajectories during active cyber-warfare scenarios.

To illustrate the behavior of a dual exposed WSN epidemic baseline prior to strategic game-theoretic control, Fig.2 (a) presents a numerical simulation of a multi-malware WSN propagation model over t $\epsilon$ $[0,1500]$ across 1000 nodes. Susceptible nodes S(t) deplete sharply as multi-vector exposure builds with $E_1(t)$ peaking approximately at $200$ around t =400. Similarly, $E_2$(t) and I(t) stabilize approximately at 80-100 while recovered nodes R(t) climb monotonically towards 990. While such modules capture multi-strain dynamics, they typically treat transition parameters as fixed or non-strategic and evaluate networks over flat topologies rather than accounting for real-time adversarial adaptation across distinct hardware tiers.

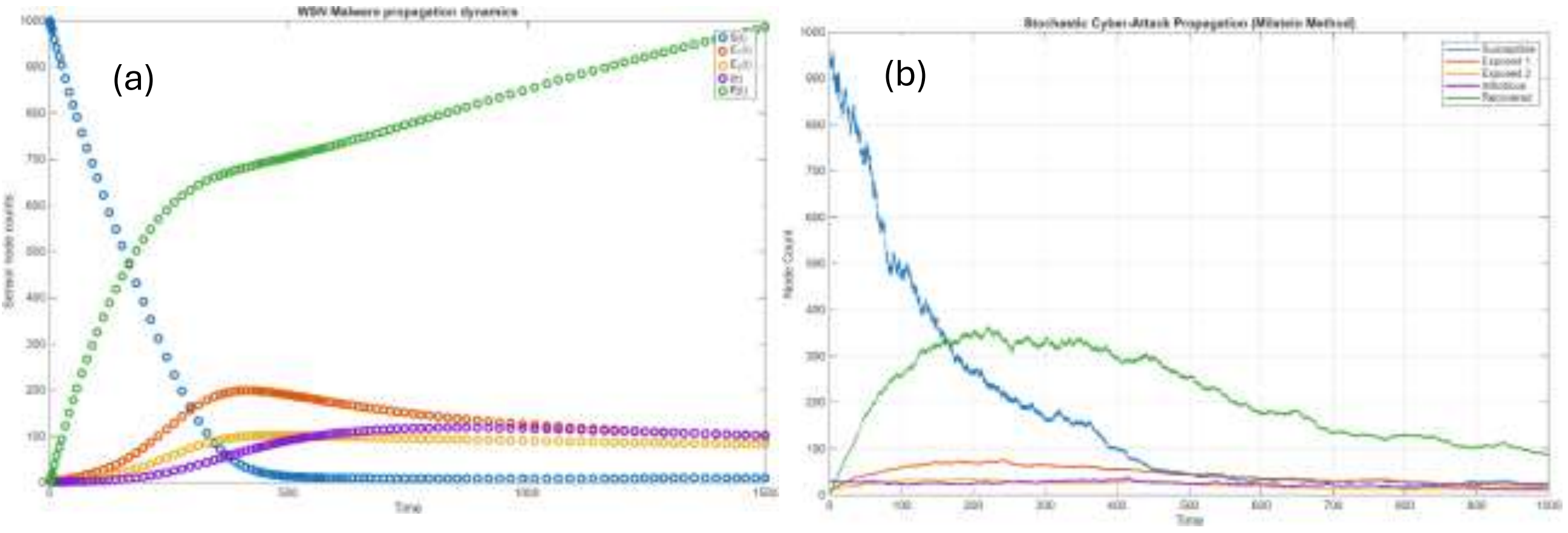

**Fig.2: Baseline and Stochastic motivation (a) Dual Exposed Wireless Sensor Network (WSN) dynamics. (b) Milstein Stochastic cyber-attack propagation.**

Stochastic differential equations (SDE) techniques, such as the Milstein numerical scheme, capture sample-path volatility and diffusion noise in cyber-attack trajectories. As shown in Fig.2 (b) over $t \in [0,1000]$ for 1000 nodes, stochastic cyber attack propagation produces pronounced sample-path fluctuations in S(t) and R(t) alongside persistent low-level Exposed ($E_1$, $E_2$) and Infectious (I) activity.

Meanwhile, Game-Theoretic and Markov Decision formulations have been explored to model strategic interaction between attackers and defenders. However, existing studies often either treat epidemic rates as static outside the game loop, omit hierarchical multi-tier WSN constraints, or leave attacker-dominant scaling regimes and solver chattering unaddressed. This paper therefore addresses the research gap across four regimes: (i) Balanced Matchup (Case I), (ii) Dynamic Stackelberg under defender-resilient scaling $\eta \geq \mu$ (Case II Baseline), (iii)Damped Dynamic Stackelberg under attacker-dominant scaling $\mu > \eta$ (Case II, Variant II), and (iv) Aggressive Quarantine (Case III).

## *III. Proposed GTI-mSEMP framework*

To capture the complex, time-variant dynamics of advanced persistent threats (APTs) navigating layered network topologies, the proposed GTI-mSEMP framework is introduced. Unlike the classical epidemiological models that assume uniform vulnerability across all interconnected devices, the GTI-mSEMP model maps malware transitions across a strictly partitioned three tier architectural hierarchy as described below [24,25]:

(i) *Gateway Layer (Tier-1):* It is the primary boundary nodes interfacing with the external public networks. This layer possesses the highest computational bandwidth and power capacity but faces aggressive, resource intensive zero-day exploit attempts targeting centralized control mechanisms.

(ii) *Routing Layer (Tier-2):* It is the intermediate infrastructure nodes for data aggregation, internal traffic routing, and state-forwarding. Vulnerabilities in this layer involve targeted exploit propagation designed to isolate local node sub-clusters.

(iii) *Peripheral Layer (Tier-*3): It includes edge devices, distributed sensors, and actuator nodes executing localized data collection, these nodes are characteristically energy constrained and highly susceptible to rapid brute-force automated malware attacks.

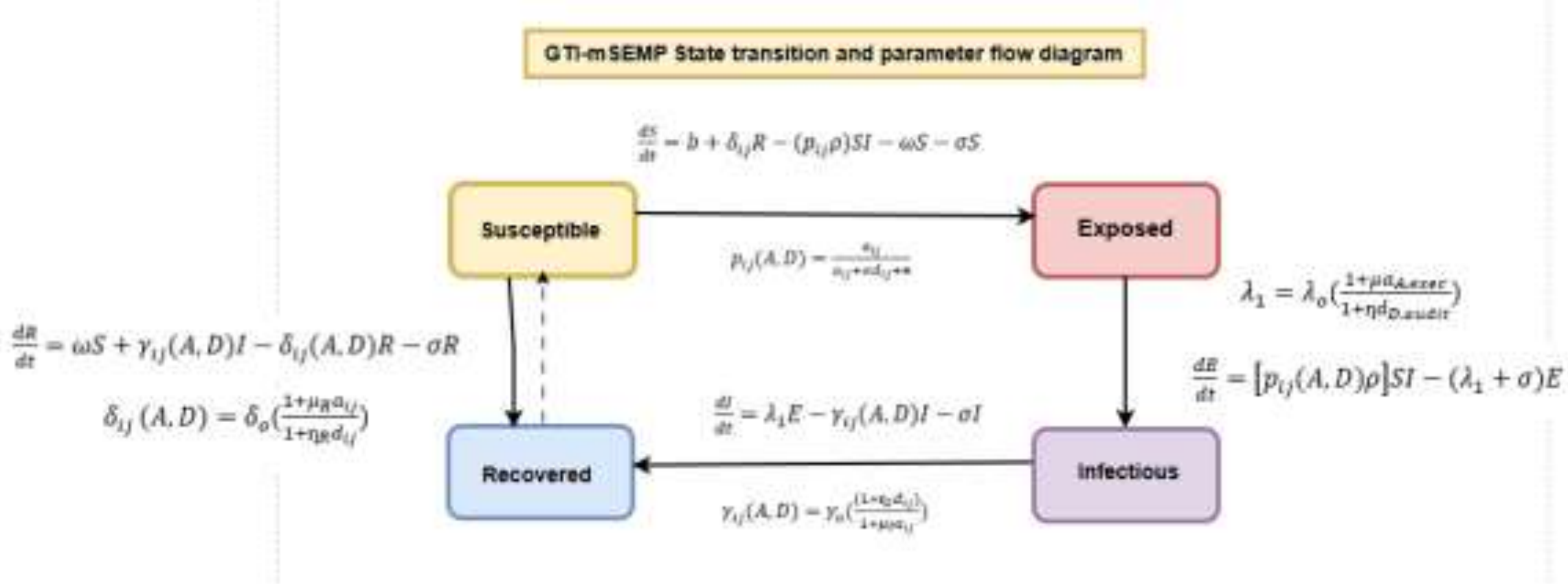

**Fig.3 : State transition compartment model for the proposed GTI-mSEMP framework**

Fig.3 shows the state transition compartment model for the GTI-mSEMP framework and highlights how the malware propagates between states along with the mathematical parameters governing dynamic transitions. It illustrates how the node population flows across *Susceptible (S)*, *Exposed (E), Infectious (I)* and *Recovered (R)* states. Transition coefficients include transition probability $(p_{ij})$, latent transition rate coefficient $(\lambda_1)$, recovery rate $\gamma_{ij}(A, D)$ and re-susceptibility coefficient $(\delta_{ij})$ are dynamically parameterized by the real-time attacker strategy matrix (A) and defender resource matrix (D).

To ensure mathematical traceability and isolate dynamic interactions between strategic adversary and defender, the proposed framework operates under the following foundational assumptions:

(i) *Constant Total Population (N):* The total number of nodes across the network layer remain fixed throughout the transient simulation window.

(ii) *Homogeneous Mixing within Layers:* Although the model separates the network into distinct tiers (Gateway, Routing and Peripheral layer), but it assumes that within any single specific layer, nodes mix homogeneously.

(iii) *Hierarchical Traffic Dependency:* The model assumes a strict hierarchical flow, where the security posture from Gateway layer constraints the operational risk profile of the Routing and Peripheral layers respectively.

(iv) *Uniform Node Capacity per Tier:* Nodes within the same tier are assumed to share identical resource constraints, computational bandwidth and power profiles.

(v) *Zero Sum Resource Constraint*: The total capital budget allocated for defense is finite.

(vi) *Rational Evasive Adversary*: The attackers are assumed to behave as rational agents with predictive awareness of the defender's resource distribution, allowing them to instantly shift their exploit capital to the network's weaker defensive gaps.

### *A. Susceptible State*

In the foundational state, the operational wireless sensor nodes have not been exposed to the malicious payload but possess inherent vulnerabilities that render them susceptible to potential exploits. The probability of a susceptible node transitioning to the latent *Exposed (E)* state within a discrete time epoch is denoted by $p$, where $p \in [0,1]$. Mathematically, $p$ is not a static scalar, but a dynamic conditional probability distribution formulated as a function of the network's spatial node density (or clustering coefficient), the attacker's operational intensity, and the defender's responsive security policy and is given by the equation (1):

$$p = f(\rho, a_A, d_D) \tag{1}$$

where, $\rho$ is the spatial density of the Wireless Sensor Network (WSN) deployment layer, $a_A \in$ A signifies the offensive strategy chosen from the attacker's action space, and $d_D \in$ D denotes the dynamic mitigation protocol executed from the defender's policy repository.

To capture the multi-layered nature of advanced persistent threats (APT) in WSN ecosystem, the offensive strategy $a_A$ is used. It is the strategy action matrix as shown in equation (2), which determines the offensive resource or energy to allocate across different target choices and attack vectors at time step, t.

$$A = \begin{pmatrix} a_{11} & \cdots & a_{1m} \\ \vdots & \ddots & \vdots \\ a_{n1} & \cdots & a_{nm} \end{pmatrix} \tag{2}$$

where $a_{ij}$ is the specific offensive investment or scan intensity directed at Node class $i$ using Exploit vector $j$.

To counter the multi-layered operational profiles of APTs, the defender's strategy profile $d_D$, can be formalized as *n x m* matrix that matches the dimension of the attacker's matrix.

$$D = \begin{pmatrix} d_{11} & \cdots & d_{1m} \\ \vdots & \ddots & \vdots \\ d_{n1} & \cdots & d_{nm} \end{pmatrix} \qquad (3)$$

where $d_{ij}$ represents the percentage of security assets, defensive compute cycles, or network bandwidth that the defender allocates to protect Node class $i$ against Exploit layer $j$.

The transition probability, $p_{ij}$can be modelled as a ratio shown in equation (4). For simplicity, we are keeping the node density, $\rho$ constant.

$$p_{ij}(A, D) = \frac{a_{ij}}{a_{ij} + \alpha d_{ij} + \in} \qquad (4)$$

where, $\alpha$ is the defensive efficiency coefficient (the effectiveness of the defensive tool) and $\in$ is the baseline network vulnerability when no defensive actions are taken.

The rate of change of Susceptible (S) stage which incorporates both the offensive and defensive strategy is given by the following equation (5).

$$\frac{dS}{dt} = b + \delta_{ij} R - (p_{ij}\rho) SI - \omega S - \sigma S \qquad (5)$$

where, b denotes the new node provisioning or join rate, $\delta$ represents re-susceptibility coefficient characterizing recovered nodes that revert to a vulnerable state due to patch expiration or configuration resets. The parameter $p_{ij}$ defines the conditional transition probability of a susceptible node entering the latent state as shown in equation (5).

The $\omega$ signifies the proactive immunization rate driven by administrative security mechanisms (for example, automated anti-virus deployment), while $\sigma$ denotes the hardware attrition rate accounting for sensor node failures induced by operational energy and battery constraints.

### B. *Exposed State (E)*

Similarly, the exposed state E is given by equation (6).

$$\frac{dE}{dt} = [p_{ij}(A, D)\rho] SI - (\lambda_{ij} + \sigma) E \qquad (6)$$

where, $[p_{ij}(A, D)\rho]SI$ denotes active transmission rate of malware propagation from Susceptible (S) to the latent Exposed (E) state. The $(\lambda_{ij} + \sigma)E$ dictates the cumulative exit rate from the latent state. The $\lambda_{ij}$ represents the transition rate at which an *Exposed node* (E) transitions to *Infectious node* (I).

In classic epidemic models, $\lambda_{ij}$ (the incubation or transition rate from Exposed to Infected) is a fixed hardware or software constant representing how long a virus naturally takes to execute. However, in an advanced cyber-warfare scenario, this transition window is actively manipulated by the defender-attacker parties involved.

The latent transition rate coefficient is bounded such that $\lambda_{ij} \in [0,1]$ within the discrete -time execution environment. This coefficient characterizes the operational latency of the malware payload: a value of $\lambda_{ij}$=0 implies absolute execution suppression while $\lambda_{ij}$=1 denotes instantaneous, single epoch initialization of the malicious process thread moving from latent *Exposed (E)* compartment to the active *Infectious (I)* compartment.

Mathematically, the latent transition coefficient $\lambda_{ij} = f(\rho, a_A, d_D)$ can be written as equation (7).

$$\lambda_{ij} = \lambda_o(\frac{1+\mu a_{A.exec}}{1+\eta d_{D.audit}}) \quad (7)$$

where, $\lambda_o$ is the baseline hardware execution speed of the malware binary, $a_{A.exec}$ is the attacker's strategic investment in payload optimization, $d_{D.audit}$ is the defender's runtime auditing intensity in that specific node layer.

If the attacker's execution acceleration coefficient, $\mu$ is large, then the malware is highly optimized for the target architecture allowing small tactical investments to drastically accelerate to outbreak speed. Similarly, the defender's detection efficiency coefficient, η is a measure of how effective the defender's auditing tools are at actively discovering, slowing down or sandboxing an unauthorized process thread.

If there are separate simultaneous malware attacks in the network, then the exposed state can be represented by $E_n$ and the corresponding transition rate as $\lambda_n$, where n$\in \{1,2,3 \ldots \ldots N\}$.

### C. *Infectious State (I)*

During the Infectious (I) state, compromised sensor nodes possess active, executing malware payloads and propagate malicious packets to neighboring susceptible and latent nodes across the WSN topology. The effective recovery rate, $\gamma_{ij}(A, D)$ is formalized as a joint function of the attacker-defender strategy profiles. From the infectious compartment, a node can transition into either pathway: (i) *Recovered pathways via administrative patch deployment and mitigation vectors*, (ii) *Permanently exits the network topology due to operational crashing or hardware malfunction induced by malicious resource exhaustion*.

Mathematically, the infectious state (I), is given by equation (8):

$$\frac{dI}{dt} = \lambda_{ij}E - \gamma_{ij}(A, D)I - \sigma I \quad (8)$$

where, $\gamma_{ij}(A, D)I$is the strategic outflow rate of nodes being successfully patched and moved to recovered state.

However, for simultaneous malware attacks, the inflow of latent nodes initializing their payloads is given by the following equation (9):

$$\text{inflow of latent nodes initializing their payloads} = \sum_{n=1}^{N} \lambda_n E_n \quad (9)$$

The strategic recovery rate includes the dynamic attacker-defender strategy profiles as similar to our susceptible probability. Mathematically, $\gamma_{ij} = f(\rho, a_A, d_D)$ can be expressed as equation (10):

$$\gamma_{ij}(A, D) = \gamma_o\left(\frac{(1+\eta_I d_{ij})}{1+\mu_I a_{ij}}\right) \qquad (10)$$

where, $\gamma_o$ is the baseline network recovery rate that could include automated backup or reboot frequency. Similarly, $\eta_I$ and $\mu_I$ are the scaling sensitivity coefficients that define the operational effectiveness of each player's active strategy during the recovery phase.

### *D. Recovered State (R)*

This encompasses the sub-population of sensor nodes that have been successfully remediated and immunized against the active malware strains. Nodes enter this state through two distinct channels: (i) proactively from the *Susceptible (S) pool* via administrative security broadcasting, or (ii) reactively from the Infectious (I) pool following successful local firmware patching ($\gamma_{ij}I$).

The rate of re-susceptibility, wherein a recovered node sheds its immunity and reverts to the Susceptible (S) state is governed by the strategic function, $\delta_{ij}$ *(A, D)*. This functional formulation captures the real-time competitive friction between the adversaries, $a_{ij}$ and $d_{ij}$.

Mathematically, the Recovered state is given by equation (11):

$$\frac{dR}{dt} = \omega S + \gamma_{ij}(A, D)I - \delta_{ij}(A, D)R - \sigma R \qquad (11)$$

where, $\omega S$ is the proactive background defense channel and $\sigma R$ is the background mortality rate.

The game dependent re-susceptibility coefficient, $\delta_{ij} = f(\rho, a_A, d_D)$is given by equation (12):

$$\delta_{ij}(A, D) = \delta_o\left(\frac{1+\mu_R a_{ij}}{1+\eta_R d_{ij}}\right) \qquad (12)$$

where, $\delta_o$ is the baseline natural rate of patch degradation and $\mu_R$, $\eta_R$ is the scaling sensitivity coefficients.

## *IV. Analytical Derivation of the Basic Reproduction Number*

To derive the analytical condition for malware eradication or system outbreak within the multi-tiered network, the basic reproduction number ($R_o$) uses the Next Generation Matrix (NGM) method. In the proposed GTI-mSEMP model, the disease state vector x, consists of the latent *Exposed ($E_i$)* and *Infectious ($I_i$)* node compartments across the heterogeneous topology tiers $i \in \{1, 2, 3\}$ (Gateway, Routing and Peripheral layers).

$$x = [\,E_1\;\; E_2\; E_3\; I_1\;\; I_2\;\; I_3]^T \qquad (13)$$

At the Disease-Free Equilibrium (DFE), no active malware footprint exists in the network.

$$x_o = [0\;0\;0\;0\;0\;0]^T \text{ and } S_i^* > 0,\ R_i^* = 0\ \forall\, i \in \{1,2,3\} \qquad (14)$$

The rate of new infections in compartment k is given by $F_k$ and the net transfer rate out of compartment k via transitions, patching or hardware attrition is represented by $V_k$.

From equation (6) and (8), the infectious state and recovered state can be expressed as equation (15) and (16).

$$\frac{dE_i}{dt} = F_{Ei} - V_{Ei} \quad (15)$$

$$\frac{dI_i}{dt} = F_{Ii} - V_{Ii} \quad (16)$$

where,

$$F_{Ei} = [p_{ij}(A, D)\rho] S^*{}_i I_i$$

$$V_{Ei} = (\lambda_{1,i}(A, D) + \sigma) E_i$$

$$F_{Ii} = 0$$

$$V_{Ii} = \lambda_{1,i}(A, D) E_i - \gamma_{ij}(A, D) I_i - \sigma I_i$$

Evaluating the Jacobians of $F$ and $V$ with respect to the infected state vector $x$ at DFE yields the linearized transmission matrix as shown in equation (17) and (18) .

$$F = [\frac{dF_k}{dx_m}]_{x_o} = \begin{pmatrix} 0_{3x3} & B_{3x3} \\ 0_{3x3} & 0_{3x3} \end{pmatrix} \quad (17)$$

$$V = [\frac{dV_k}{dx_m}]_{x_o} = \begin{pmatrix} \Lambda_{3x3} & 0_{3x3} \\ -L_{3x3} & \Gamma_{3x3} \end{pmatrix} \quad (18)$$

where,

$$B_{ii} = p_{ij}(A, D)\rho S^*{}_i$$

$$\Lambda_{ii} = \lambda_{1,i}(A, D) + \sigma$$

$$L_{ii} = \lambda_{1,i}(A, D)$$

$$\Gamma_{ii} = \gamma_{ij}(A, D) + \sigma$$

Computing reproduction number ($R_o$) using the Next Generation Matrix is shown in equation (21).

$$R_o = \rho(FV^{-1}) \quad (19)$$

$$R_o = \frac{[p_{ij}(A,D)\rho S^*{}_i].[\lambda_{1,i}(A,D)]}{[\lambda_{1,i}(A,D)+\sigma].[\gamma_{ij}(A,D)+\sigma]} \quad (20)$$

Similarly,

$$R_o = \frac{\left[\left(\frac{a_{ij}}{a_{ij}+\alpha d_{ij}+\in}\right)\rho S^*{}_i\right] \cdot [\lambda_o(\frac{1+\mu a_{ij}}{1+\eta d_{ij}}]}{\left[(\lambda_o\left(\frac{1+\mu a_{ij}}{1+\eta d_{ij}}\right)+\sigma\right] \cdot [\gamma_o\left(\frac{1+\eta_I d_{ij}}{1+\mu_I a_{ij}}\right)+\sigma]} \tag{21}$$

If $R_o$ (A, D) < 1, the disease-free equilibrium (DFE) is asymptotically stable (dynamic defense policy successfully contains and extinguishes the malware propagation). Conversely, $R_o$ (A, D) > 1 leads to an epidemic outbreak.

### *V. Proof of Equilibrium States and Game Theoretic Analysis*

Mathematically the conditions under which the malware persists or eradicated in a multi-tier network to establish system's equilibrium points are established in this section. The disease-free equilibrium ($E_o$) represents a state where there is no active or latent malware that exists across any network layer. Mathematically it can be represented as:

$$E_o = [S_1^* \ S_2^* \ S_3^* \ 0 \ \ 0 \ 0 \ 0 \ 0 \ 0 \ R_1^* \ R_2^* \ R_3^*]^{\mathrm{T}} \tag{22}$$

Hence equations (1) and (11) can be written as:

$$S_i^* = \frac{b_i}{\omega_i+\sigma_i}$$

$$R_i^* = \frac{\omega_i S_i^*}{\delta_i + \sigma_i}$$

*Theorem-1: Determining the Local Stability of Disease-Free Equilibrium (DFE) via Routh-Hurwitz*

The rates of change for Exposed ($E_i$) and Infectious ($I_i$) compartments in layer *i* are given by:

*Exposed Compartment* ($E_i$)

$$\frac{dE_i}{dt} = g_1(E_i, I_i)$$

*Infectious Compartment* ($I_i$)

$$\frac{dI_i}{dt} = g_2(E_i, I_i)$$

The general Jacobian matrix, $J_i$ for the infected state variables $[E_i, I_i]^{\mathrm{T}}$ is the matrix of partial derivatives:

$$J_i = \begin{pmatrix} \frac{dg_1}{dE_i} & \frac{dg_1}{dI_i} \\ \frac{dg_2}{dE_i} & \frac{dg_2}{dI_i} \end{pmatrix} \tag{23}$$

At the DFE,

$$J_i(E_o) = \begin{bmatrix} -(\lambda_{1,i} + \sigma) & p_{ij}\rho S_i^* \\ \lambda_{1,i} & -(\gamma_{ij} + \sigma) \end{bmatrix} \quad (24)$$

The characteristic equation is given by equation (25)

$$\det(J_i(E_0) - kI_{2x2}) = 0 \quad (25)$$

where k represents eigenvalues and $I_{2x2}$ is the identity matrix.

$$\det \begin{bmatrix} -(\lambda_{1,i} + \sigma) - k & p_{ij}(A,D)\rho S^* \\ \lambda_{1,i} & -(\gamma_{ij} + \sigma) - k \end{bmatrix} = 0$$

Solving equation (25)

$$[(-(\lambda_{1,i} + \sigma) - k). -(\gamma_{ij} + \sigma) - k]\text{-}[\lambda_{1,i}. p_{ij}(A,D)\rho S^*] = 0$$

$$k^2 + k[(\lambda_{1,i} + \sigma) + (\gamma_{ij} + \sigma)] - \lambda_{1,i}. p_{ij}(A,D)\rho S^* + [(\lambda_{1,i} + \sigma).((\gamma_{ij} + \sigma)] = 0 \quad (26)$$

Comparing to a second order polynomial equation:

$$k^2 + A_1 k + A_0 = 0 \quad (27)$$

The Routh-Hurwitz equation states that all roots have negative real parts (Re (k) < 0) if all coefficient $A_1 > 0 \; and \; A_0 > 0$.

Hence, equating coefficient (26) and (27).

$$A_1 = \lambda_{1,i} + \gamma_{ij}(A,D) + 2\sigma$$

Since execution rate ($\lambda_{1,i}$), recovery rate ($\gamma_{ij}$) and attrition rate ($\sigma$) are positive physical parameters, therefore $A_1 > 0$.

Similarly,

$$A_0 = -\lambda_{1,i}. p_{ij}(A,D)\rho S^* + [(\lambda_{1,i} + \sigma).((\gamma_{ij} + \sigma)]$$

$$[(\lambda_{1,i} + \sigma).(\gamma_{ij} + \sigma)] > \lambda_{1,i}. p_{ij}(A,D)\rho S^*$$

$$1 > \frac{\lambda_{1,i}. p_{ij}(A,D)\rho S^*}{[(\lambda_{1,i} + \sigma).(\gamma_{ij} + \sigma)]}$$

Compared with equation (20)

$$R_0(A,D) < 1$$

This proves that all eigenvalues of $J_i(E_0)$ has negative parts and $E_0$ is locally asymptotically stable.

*A. Theorem- 2: Existences and Uniqueness of Sub-game Perfect Stackelberg Equilibrium*

The multi-tier network game $\Gamma = (A, D, R_o)$ can be defined over the compact and convex strategy spaces is given by equation (28):

$$A = \{\boldsymbol{A} \in \mathbb{R}^{3X2} | \sum_{i=1}^{3}\sum_{j=1}^{2} c_{A,ij}a_{ij} \leq K_A, a_{ij} > 0\} \quad (28)$$

where, A is the feasible attack strategy matrix of size 3(network tier) x 2 (attack vector: Brute Force and Zero-Day), $c_{A,ij}$ is the cost coefficient for execution, $a_{ij}$ is the intensity or energy allocated to attack vector j on network tier, i and $K_A$ is the attacker's maximum available budget.

Similarly,

$$D = \{D \in \mathbb{R}^{3X2} | \sum_{i=1}^{3}\sum_{j=1}^{2} c_{D,ij}d_{ij} \leq K_D, d_{ij}^{min} \leq d_{ij} \leq d_{ij}^{max}\} \quad (29)$$

where, $K_A, K_D > 0$ are fixed execution budgets, and $R_0(A, D)$ is the basic reproduction number derived earlier in the section.

Then:

*Existence:* There exists at least one Subgame-Perfect Stackelberg Equilibrium $(A^*, D^*) \in AxD.$

*Uniqueness:* If $R_0(A, D)$ is strictly concave in A for any given D and strictly convex in D along the optimal follower path $A^*(D)$, the equilibrium $(A^*, D^*)$ is unique.

*B. Theorem-3: Mathematical Proof of the Existence of the Subgame-Perfect Stackelberg Equilibrium*

The game $\Gamma_{stack} = (A, D, R_o)$ admits at least one subgame -perfect Stackelberg Equilibrium $(A^{SE}, D^{SE}) \in AxD$. Furthermore if $p_{ij}(a_{ij}, d_{ij})$ is strictly concave in $a_{ij}$ and strictly convex in $d_{ij}$, the equilibrium strategy pair $(A^{SE}, D^{SE})$. Backward induction allows to find the optimal choice in game theory to find the Subgame-Perfect Equilibrium.

*Leader Strategy (Stage -I):* Substituting the continuous reaction function $A^*(D)$ into the leader's objective yields the reduced function $f(D) = R_o(A^*(D), D)$. Since $R_o$ and $A^*(D)$ are continuous, $f(D)$ is continuous over the compact and convex set D. Applying the Weierstrass's Extreme Value Theorem (which states that every continuous real valued function on a close interval can be uniformly approximated by a polynomial) guarantees the existence of an optimal defense strategy $D^{SE} \epsilon \arg min_{\boldsymbol{D} \epsilon D} f(D)$.

*Follower subgame (Stage -II)*:

For a fixed defensive posture $\boldsymbol{D} \epsilon D$, the attacker solves the follower problem:

$$max_{\boldsymbol{A} \in A} R_o(A, D)$$

The constraint set A is defined by the linear inequalities $\sum_{i,j} c_{A,ij} \leq K_A$ and $a_{ij} \geq 0$.

Thus, space **A** is a closed, bound and convex subset of $\mathbb{R}^6$. Similarly, **D** is a compact subset of $\mathbb{R}^6$. The spectral radius function $R_o(A, D) = \rho(FV^{-1})$ is a continuous function ($\mathcal{C}^0$) over $A\ xD$, as matrix entries in $F$ and $V$ depend continuously on infection or transition probabilities $p_{ij}(a_{ij}, d_{ij})$.

By Weierstrass's Extreme Value Theorem (which states that is a real valued function is continuous on the closed and bounded interval, then it must attain a maximum and a minimum at least once), a maximum $A^*(D)$ exists for every $\boldsymbol{D} \in D$. Strictly concavity guarantees that the optimal reaction mapping $A^*(D) = \mathcal{R}_{\boldsymbol{A}}(D)$is single valued and continuous (via Berge's Maximum Theorem).

$$A^*(D) = argmax_{\ \boldsymbol{A} \in A} R_0(A, D)$$

is nonempty and compact for every $\boldsymbol{D} \in D$.

*Leader's Optimization problem:*

The defender solves:

$$min_{\boldsymbol{D} \in D}\ \ R_o^*(D)$$

Since $R_o^*(D)$ is a continuous on the compact domain D, applying Weierstrass's Extreme Value theorem a second time guarantees that the leader's optimization problem attains its minimum at $D^* \in D$.

Selecting any $A^* \in A^*(D^*)$completes the proof of existence for the SPSE pair $(A^*, D^*)$.

*Equilibrium pair:* The strategy pair $(A^{SE}, D^{SE})$ where $A^{SE}$=$A^*(D^{SE})$, constitutes a Subgame-Perfect Stackelberg Equilibrium (SPSE) satisfying:

$R_o(A^{SE}, D^{SE}) \leq R_o(R_A(D), D),\ \forall\ \boldsymbol{D} \in D.$

## *VI. Case Studies for Multi-Tier Architectural Hierarchy*

To evaluate the operational dynamics of the proposed GTI-mSEMP framework, the baseline simulation model categorizes the network topology into three distinct heterogeneous node classes: *(i) Core Gateway, (ii) Intermediate Routing Nodes, and (iii) Low-Power Peripheral Sensors.* Correspondingly, two primary adversarial vectors are evaluated to represent varying tiers of threat severity*:* (i) *Brute-Force Attack (BFA)*, characterized by automated credential guessing routines, and (ii) Zero-Day Exploits, representing sophisticated attacks targeting undocumented software or hardware vulnerabilities [26-29].

The strategic resource allocation profiles of both the attacker and defender are mapped to a joint 3 x 2 game matrix (A, D ∈ $R^{3\,x2}$), where the row vectors correspond to the targeted topological node layers and column vectors quantify the relative capital assigned to each exploit scenario. To analyze the systemic sensitivity under varying tactical conditions, three foundational operational regimes are investigated: *Case I- Balanced Equilibrium Matchup* (which establishes a baseline symmetric resource allocation between both players), *Case II- Evasive Exploit Surge* (simulating a strategic mismatch where the defender is severely outmaneuvered), *Case III- Aggressive Quarantine* (characterizing a hardened network posture reinforced by enhanced administrative defense controls).

To mathematically, evaluate the three strategic regimes, the resource allocation matrices for the Attacker (A) and Defender (D) are parameterized across the network layers (Row 1: Gateways, Row 2: Routing Nodes, Row 3: Peripherals) and exploit vectors (Column 1: BFA and Column 2: Zero-Day Attack).

*Case I: Balanced Equilibrium Matchup (Symmetric Baseline)*

Case I models a benchmark operational state where the adversary and defender possess symmetric information, equivalent operational capital, and balanced strategic priorities across the multi-tier hierarchy. Neither player holds an asymmetric technical advantage over the opponent with an unmitigated exploit vector.

$$A_1 = \begin{pmatrix} 0.40 & 0.20 \\ 0.30 & 0.10 \\ 0.00 & 0.00 \end{pmatrix}$$

$$D_1 = \begin{pmatrix} 0.40 & 0.20 \\ 0.30 & 0.10 \\ 0.00 & 0.00 \end{pmatrix}$$

Gateway Layer (40% BFA and 20% Zero-Day Attack): Both players assign 60% of their total capital to the Gateway Layer tier. The adversary recognizes that compromising public facing gateways provides centralized control and high bandwidth propagation opportunities. The defender systematically counters by deploying 60% of their security budget (such as firewalls, deep packet inspection, hardware authenticity) directly to the gateway perimeter.

Routing Layer (30% BFA and 10% Zero-Day Attack): The intermediate routing layer receives 40% of the budget. Attackers attempt to compromise data-forwarding nodes to isolate sub-clusters, while defenders deploy traffic auditing and rate-limiting protocols proportionally to maintain routing integrity.

Peripheral Layer (0% Direct Investment): Peripheral edge sensors receive zero targeted capital allocation under this baseline because their individual failure does not immediately compromise network infrastructure and their high node density relies on baseline automated security controls rather than dedicated budget expenditure.

*Case II: Dynamic Evasive Exploit Surge (Stackelberg Game Equilibrium)*

Case II simulates a dynamic, highly adversarial environment where a rational attacker actively predicts and exploits residual defensive vulnerabilities. Rather than maintaining a static distribution, the defender (Leader) and attacker (Follower) continuously update their strategy matrices $D^{*}$(t) and $A^{*}$(t) at each time epoch via the KKT Stackelberg optimization engine ($min_D max_A R_o(A, D)$).

$$A_2 = \begin{pmatrix} 0.00 & 0.10 \\ 0.00 & 0.90 \\ 0.00 & 0.00 \end{pmatrix}$$

$$D_2 = \begin{pmatrix} 0.50 & 0.10 \\ 0.30 & 0.10 \\ 0.00 & 0.00 \end{pmatrix}$$

Attacker Strategic Evasion (90% Routing Zero-Day Surge): The adversary detects that the defender has heavily hardened the Gateway Layer against credential -guessing attacks ($d_{11} = 0.50$). To outmaneuver this perimeter defense, the attacker shifts 90% of their offensive capital into a sophisticated Zero-Day exploit targeted specifically at internal routing nodes ($a_{22} = 0.90$), by passing gateway firewalls entirely.

Defender Mismatch and Reactive Adaption: The initial defensive posture ($D_2$) leaves the routing layer under-protected against Zero-Day threats ($d_{22} = 0.10$). This strategic asymmetry induces an immediate epidemic spike, forcing the closed-loop game engine to dynamically shift defensive capital from the Gateway perimeter toward internal routing node auditing ($d_{22}^{*}(t)$ ↑)to reestablish KKT stationarity.

*Case III: Aggressive Quarantine*

Case III models a perimeter -centric heavily hardened security policy where administrative controls enforce aggressive isolation protocols at primary entrance points. The defender prioritizes absolute gateway containment to eliminate external intrusion attempts before payloads can reach internal nodes.

$$A_3 = \begin{pmatrix} 0.50 & 0.20 \\ 0.20 & 0.10 \\ 0.00 & 0.00 \end{pmatrix}$$

$$D_3 = \begin{pmatrix} 0.80 & 0.10 \\ 0.10 & 0.00 \\ 0.00 & 0.00 \end{pmatrix}$$

Asymmetric Perimeter Hardening (80% Gateway BFA Defense): The defender allocates an aggressive 80% of their security budget strictly to BFA mitigation at the Gateway tier ($d_{11} = 0.80$). This creates a “hardened shell” that drastically suppresses the transmission probability $p_{11}$(A, D)., rapidly driving down the initial infection rate among boundary nodes.

Vulnerability via Defense Concentration: Because 90% of total defensive capital is locked at Tier 1 ($d_{11} + d_{12} = 0.90$). the intermediate routing nodes ($d_{21} = 0.10, d_{22} = 0.00$) and peripheral devices ($d_{31} = 0.00, d_{32} = 0.00$) are left essentially unhardened. If an adversary successfully penetrates the perimeter or initiates an insider attack at Tier 2 or 3, the lack of internal defense causes rapid node depletion.

### *VII. Simulation Setup, Numerical Results and Discussion*

Unlike static baseline allocations, Case II models the real-time competitive interactions between a proactive Defender (Leader) and an adaptive, evasive Attacker (Follower). At each discrete numerical time step $t_k$, the state profile $x(t_k) = [S_i(t_k), E_i(t_k), I_i(t_k), R_i(t_k)]^T$ is evaluated across all tiers $i \in \{1,2,3\}$. The strategic allocation profiles $A^*(t_k)$ and $D^*(t_k)$ are determined by solving subgame-perfect Stackelberg Equilibrium problem:

$$D^*(t_k) = argmin_{\boldsymbol{D} \in D}[max_{\boldsymbol{A} \in A} R_0(A, D \mid x(t_k))] \quad (30)$$

$$A^*(t_k) = argmax_{\boldsymbol{A} \in A} R_0(A, D^* \mid x(t_k)) \quad (31)$$

subject to $\sum_{i=1}^{3} \sum_{j=1}^{2} c_{D,ij}\, d_{ij} \leq K_D, \sum_{i=1}^{3} \sum_{j=1}^{2} c_{A,ij}\, a_{ij} \leq K_A$

The calculated decision matrices $A^*(t_k)$ and $D^*(t_k)$ update the state transition parameters $p_{ij}(t_k), \lambda_{1,i}(t_k)$, $\gamma_{ij}(t_k)$ and $\delta_{ij}(t_k)$ for the subsequent 4$^{th}$ Order Runge- Kutta (RK4) integration epochs. This guarantees that the trajectory tracking directly reflects the optimal KKT stationary conditions derived in Section V.

To evaluate the dynamic closed-loop performance of the proposed GTI-mSEMP framework, a numerical simulation environment was implemented in MATLAB, with comparative state trajectories illustrated in Fig.3.Algorithm 1 outlines the procedure for evaluating the variation in total Susceptible (S(t)) and Recovered (R(t)) node trajectories across all three operational regimes. For Cases I and III, the static trajectory metrics $(A_1, D_1)$ and $(A_3, D_3)$ are assigned across the network tiers to model fixed baseline setups. Conversely, Case II evaluates the subgame-Perfect Stackelberg Equilibrium pair ($A^*$,$D^*$) at each step by solving equation (30) and (31).

For Cases I and III, the system of coupled non-linear differential equations is solved numerically using the 4$^{th}$ Order Runge Kutta (RK4) integration method with a fixed step size dt = 0.02 over a simulation horizon of t ∈ [0,50] epochs across a heterogeneous three-tier sensor topology. For Case II, a nested bilevel optimization function executes Sequential Quadratic Programming (SQP) at each discrete epoch to determine the optimal defender allocation ($D^*$) that minimizes the worst-case basic reproduction number $R_o$, effectively countering the adversary's attempt to exploit residual vulnerability gaps. The simulation parameters used across all cases are detailed in Table-I.

**Table I: Parameters used in the GTI-mSEMP Framework simulation**

| Parameters | Magnitude | Description |
|---|---|---|
| b | 0.500 | New node provisioning rate |
| $\omega$ | 0.050 | Proactive immunization rate |
| $\sigma$ | 0.002 | Hardware attrition rate |
| $\rho$ | 0.002 | Node density |
| $\alpha$ | 5.000 | Defensive efficiency coefficient |
| $\in$ | 0.020 | Baseline network vulnerability |
| $\lambda_o$ | 0.150 | Baseline hardware execution speed |
| $\gamma_o$ | 0.080 | Baseline network recovery rate |
| $\delta_o$ | 0.040 | Baseline natural rate of patch degradation |

**ALGORITHM 1: Multi-Regime Comparative Network Simulation (For Figure 3)**

Input : Initial state vector $X_0 = [S_i(0), E_i(0), I_i(t), R_i(t)]^T$ for network tiers i ∈ {1, 2, 3};
Time horizon t = 50 epochs;
Numerical integration step size dt = 0.02;
Static strategy allocation matrices $(A_1, D_1)$ for Case I and $(A_3, D_3)$ for Case III;
Capital budgets $K_A$= 1.0, $K_D$ = 1.0 and cost matrices $C_A$, $C_D$ for Case II.

Output : Compartment trajectories X_CaseI(t), X_CaseII(t), X_CaseIII(t) and aggregate curves S(t), R(t).

Initialize discrete time grid t = [0, dt, 2dt, ..., T]
% --- STAGE 1: NUMERICAL INTEGRATION FOR STATIC REGIMES (CASES I and III)
FOR EACH static regime M ∈ {Case I, Case III} DO
Set initial compartment state vector X_M(1) = X_0
FOR discrete time epoch k = 1 TO (T/dt - 1) DO
Extract current state profile X_M(k)
Compute transition probability $p_{ij}$(A_M, D_M) via equation (4)
Compute payload latency $\lambda_{ij}${1,i}$(A_M, D_M)$ and recovery rate $\gamma_{ij}(A_M, D_M)$ via equations (7), (10).
Evaluate non-linear ODE system slopes f(X_M(k), $A_M$, $D_M$) via equations (5), (6), (8), (11)
Advance state vector X_M(k+1) using 4th-Order Runge-Kutta (RK4) integration step
END FOR
END FOR

% --- STAGE 2: DYNAMIC STACKELBERG OPTIMIZATION LOOP (CASE II) ---
Set initial state vector X_CaseII(1) = X_0 and initial strategy guesses A^(0), D^(0)
FOR discrete time epoch k = 1 TO (T/dt - 1) DO
Extract local compartment state X_CaseII(k) across Gateway, Routing, and Peripheral tiers

% Solve Leader-Follower Bilevel Optimization Problem
D*(k) = argmin_{D ∈ D} [ max_{A ∈ A} R_0(A, D | X_CaseII(k)) ]
A*(k) = argmax_{A ∈ A} R_0(A, D*(k) | X_CaseII(k))
Subject to: ∑ ∑ c_{$D_{ij}$} * $d_{ij}$≤ K_D and ∑ ∑ c_{$A_{ij}$} * $a_{ij}$≤ K_A

Update dynamic transition parameters $p_{ij}$ (k), $\lambda_{ij}${1,i}(k), $\gamma_{ij}$ (k) using optimal matrices A*(k), D*(k)
Advance state vector X_CaseII(k+1) over interval [t_k, t_k + dt] via RK4 integration step
END FOR

% --- STAGE 3: AGGREGATE POPULATION SUMMATION ---
FOR EACH regime M ∈ {Case I, Case II, Case III} DO
Compute total susceptible nodes : S_M(t) = ∑_{i=1}^3 S_i,M(t)
Compute total recovered nodes : R_M(t) = ∑_{i=1}^3 R_i,M(t)
END FOR

RETURN S_CaseI(t), S_CaseII(t), S_CaseIII(t) and R_CaseI(t), R_CaseII(t), R_CaseIII(t) for plotting Figure 3.

Fig. 4(a) compares the Susceptible S(t) node trajectories across the operational regimes. It is evident that Case II experiences the highest rate of depletion, where the susceptible population declines most rapidly from 1000 nodes to approximately 255 nodes over the 50-epoch horizon. Case II's sharper drop reflects the adaptive exploit surge under the Stackelberg game formulation. The bilevel optimization engine (the Stackelberg game) calculates the optimal strategy for both players $A^*$ and $D^*$ each time by playing out the minmax/ maximin logic on the reproduction number. Once the matrices are calculated they are applied to the corresponding transition parameters such as $p_{ij}$, $\lambda_{ij}$ and $\gamma_{ij}$. The RK4 integration then updates across all the compartments simultaneously yielding a drop in Susceptible S(t) node population, temporary spikes in Exposed E(t) and Infectious I(t) and rise in Recovered R(t) node population.

Similarly, Case III exhibits a faster rate of susceptible node depletion than Case I (settling around 355 nodes versus 455 nodes) which is governed by the static resource allocation configuration in matrices $A_3$ and $D_3$ respectively.

Fig.4 (b) illustrates the corresponding cumulative recovered node trajectories R(t) across the three operational regimes over the 50-epoch horizon. While all cases exhibit a monotonic rise in recovery as nodes transition through the compartment pipeline, distinct long-term saturation behaviors emerge. Case II achieves the highest overall recovery count (converging near 570 nodes), which correlates directly with its earlier higher infection volume- forcing a larger absolute number of nodes through the remediation and patching cycle once the Dynamic Stackelberg optimization closes the loop. Conversely, Case III exhibits the lowest ultimate recovery saturation (~ 530 nodes) , because the aggressive boundary quarantine successfully blocks intrusion at the entry gate layer, fewer nodes are compromised in the first place, reducing the requisite volume of recovery operations. Case I maintains a balanced intermediate trajectory, stabilizing between the dynamic adaptation of Case II and the static perimeter isolation of Case III.

Fig.4(a) and 4(b) provides a comparative evaluation that enables SOC analysts to assess resource allocation trade-offs and decision-making performance under game-theoretic optimization. As illustrated in Fig.3(a), Case II experiences the deepest transient compromise, with 745 nodes leaving the susceptible pool, out of which the dynamic Stackelberg optimization achieves a recovery of approximately 570 nodes (yielding a 76.5% recovery rate). Conversely, Case III restricts total infection of 645 nodes through aggressive perimeter hardening, achieving a final recovery of 530 nodes (an 82.5% recovery rate). Meanwhile, Case I establishes a stable symmetric baseline where 545 nodes leave the susceptible pool and virtually all of them successfully transition through the remediation pipeline to achieve full recovery., demonstrating how dynamic adaptation in Case II and perimeter quarantine in Case III trades baseline equilibrium for specific security postures.

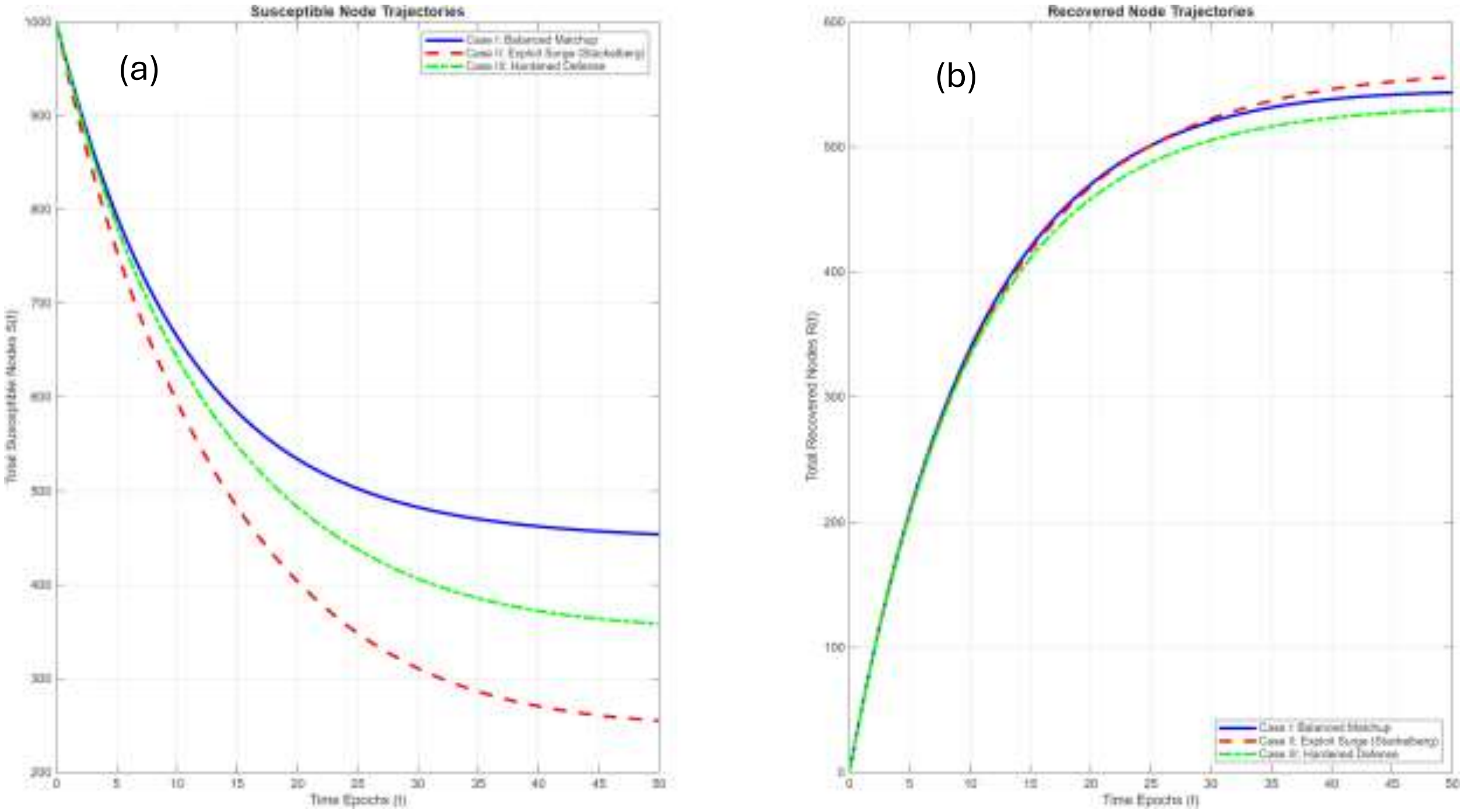

**Fig. 4:** Operational trajectories of the *Susceptible (S)* and *Recovered (R)* node populations for Case I: Balanced Matchup, Case II: Exploit Surge and Case III: Hardened Defense. *Left: Susceptible* Node Trajectories. *Right:* Recovered Node Trajectories

**Algorithm 2: Dynamic Stackelberg Game-Theoretic SEIR Integration (Case II- Fig.4)**

Input: Total horizon T_final, time step dt, initial states X_0, budgets K_A, K_D

Output: State trajectory matrix X, defensive budget history D_history

```
Initialize strategy matrices A_curr and D_curr uniformly
Set time vector t = 0 : dt : T_final
for k = 1 to length(t) - 1 do
    Extract current network state X_k = X(k, :)

    % --- Outer Leader Loop (Defender Minimization) ---
    Solve min_D max_A R_o(A, D | X_k) via fmincon (SQP)
    Update optimal defender strategy matrix D_curr

    %--- Inner Follower Loop (Attacker Best-Response) ---
    Compute best-response A_curr given D_curr and X_k
    Record tier-specific defense allocations into D_history(k, :)

    %--- Numerical Integration (4th-Order Runge-Kutta) ---
    Compute RK4 slopes (k1, k2, k3, k4) using mSEMP_ode(X_k, A_curr, D_curr)
    Advance state: X(k+1) = X_k + (dt / 6) * (k1 + 2*k2 + 2*k3 + k4)
end for
return X, D_history
```

The co-evolution of network epidemiological trajectories and adaptive security posture adjustments is depicted in Algorithm-2 and Fig. 5. The GTI-mSEMP framework is implemented via a discrete time execution loop. The architecture bridges macro-scale compartment dynamics with micro-scale game theoretical updates, reflecting real-world Security Operations Center (SOC) environments where defenders must continuously react to evolving threat landscapes.

As outlined in Algorithm 2, the simulation processes the 50 epochs horizon by coupling a nested bi-level optimization engine with the $4^{th}$ order Runge-Kutta (RK4) numerical integration. The mechanism of this integration proceeds through several distinct analytical stages:

1. Stackelberg Equilibrium and Leader Follower Dynamics: At each discrete time step k, the network state vector $X_k$ is evaluated. Acting as the Stackelberg leader, the defender computes an optimal resource allocation strategy by anticipating how an intelligent adversary will react. This is operationalized via Sequential Quadratic Programming ($f_{mincon}$), where the outer loop minimizes the worst-case basic Reproduction number ($R_o$).
2. Inner Loop Attacker Best-Response: Inside the leader's evaluation, the follower (Attacker) computes the optimal exploit distribution across network tiers ($c_A, K_A$) to maximize infection propagation. This nested structure ensures that the defense strategy is robust against worst case adversary behavior rather than static assumptions.
3. Parameter Modulation and State Transitions: The resulting strategic allocation shares $d_i^*(t)$dynamically modulate the system's effective transmission ($\beta_{ij}$) and recovery rates ($\gamma_{ij}$) across the gateway, routing and peripheral tiers.
4. Numerical State Propagation: With the updated game-theoretic parameters obtained in step k, the system state is advanced using the $4^{th}$ Order Runge-Kutta integration (mSEMP-ode). Computing four intermediate slope increments ($k_1, k_2, k_3, k_4$) to maintain high numerical stability across the transient and steady state phases.

The numerical simulations and nested Stackelberg optimizations were implemented in MATLAB. To solve the constrained non-linear leader-follower problem at discrete time step, the framework utilizes MATLAB's $f_{mincon}$ solver executing the Sequential Programming (SQP) algorithm, with objective and tolerance thresholds optimized for rapid convergence across the 50 epochs. The optimal resource allocation, ($d_i^*$) for the leader (Defender) is verified through the Karush-Kuhn-Tucker (KKT) optimization conditions across the three network tiers.

As evident in Fig. 5, the initial defensive resource budget allocation for the Gateway layer increases from 0.09 to 0.20, while the allocation for the Peripheral and Routing layer adjusts to 0.35 and 0.47 respectively, maintaining the total defender budget constraints ($K_D \approx 1.0$). This adaptive reallocation is driven by the KKT optimization framework: as the system detects diminishing susceptible nodes and early infection dynamics, it shifts its resources towards the Gateway tier to establish a robust perimeter defense. Consequently, although a minor infection peak is observed over the initial 10 epochs, the closed-loop optimization effectively suppresses runaway propagation. By dynamically reallocating defensive resources via KKT conditions, the framework successfully curtails secondary transmission, preserving a susceptible baseline while ensuring that active exposures are systematically steered towards recovery, approximately 585 nodes (representing a 99% recovery rate at the simulation horizon).

Fig. 6 demonstrates an attacker-dominant regime ($\mu > \eta$), where the attacker's exploitation capability significantly outweighs the defender's mitigation speed. The $\mu$ and $\eta$ are strategic scaling factors (or sensitivity coefficients) that control how strongly the attacker's and defender's resource allocation affects the epidemiological transition rates:

- $\mu$ (Attacker's Scaling factor): Measures the effectiveness of the attacker's exploit resources (A). Higher values of $\mu$ means the attacker accelerates malware propagation, increases transition rates to Infectious state ($\lambda_{ij}$), and slows down recovery ($\gamma_{ij}$) or drives faster immunity loss ($\delta_{ij}$).
- $\eta$ (Defender's Scaling factor): Measures the effectiveness or mitigation power of the defender's security resources (D). Higher the value of $\eta$ means that the defender strongly suppresses transmission and speeds up the recovery ($\gamma_{ij}$) and reinforces node resilience.

Therefore, when $\mu > \eta$, the network is operating in an attacker-dominant regime. Even if both sides have equal budgets ($K_A = K_D = 1$), every unit of attacker effort yields a disproportionately higher impact on spreading the exploit that the unit of defender effort yields in containing it. This represents scenarios like zero-day exploits or stealthy polymorphic malware that can outpace a standard Security Operations Center (SOC) patching spread.

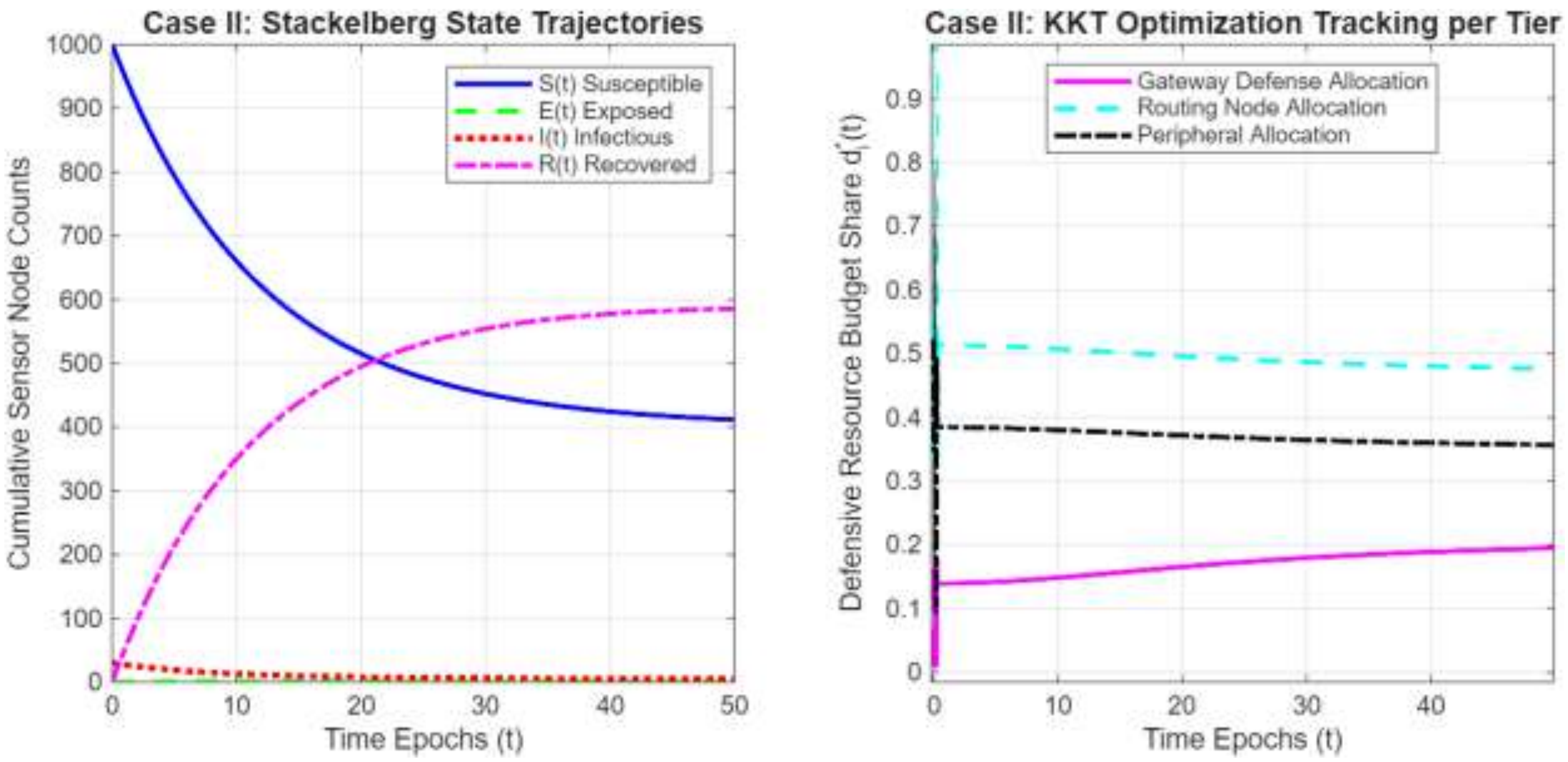

**Fig. 5:** Closed-Loop Evasive Game Analysis (Case II). *Left:* Temporal evolution of cumulative sensor node counts distributed across SEIR compartments. *Right:* Dynamic convergence of defensive resource allocation tracking across heterogeneous network layers.

Algorithm 3 outlines the computational procedure for Case II, Variant II (Fig.6). Whereas Algorithm-2 models a baseline regime in which the defender mitigation scaling outweighs or matches attacker amplification ($\eta \geq \mu$), Algorithm 3 inverts these scaling coefficients across both the reproduction objective evaluation ($R_0$) and the compartmental SEIR transition dynamics (such as $\mu_I$=5.0 > $\eta_I$=2.0 and $\mu_R$= 6.0 > $\eta_R = 1.5$).

**Algorithm 3: Damped Stackelberg Sensitivity Analysis (Case II, Variant II, Fig. 5)**

Input: Extended horizon T_final = 150, time step dt = 0.1, initial states X_0

Output: State trajectory matrix X, damped defensive budget history D_history

```
Initialize strategy matrices A_curr and D_curr uniformly
Set time vector t = 0 : dt : T_final
for k = 1 to length(t) - 1 do
    Extract current network state X_k = X(k, :)

    // --- Outer Leader Loop (Defender Minimization under Attacker Advantage) ---
    Solve min_D max_A R0_variant2(A, D | X_k) via fmincon (SQP)
    Retrieve raw optimal defender strategy D_new

    // --- Strategy Damping / Rate Limiting (Smoothing Chattering) ---
    Update D_curr = 0.7 * D_curr + 0.3 * D_new

    // --- Inner Follower Loop (Attacker Best-Response under mu > eta) ---
    Compute best-response A_new given D_curr and X_k
    Update A_curr = 0.7 * A_curr + 0.3 * A_new
    Record active tier-specific budget sums into D_history(k, :)

    // --- Numerical Integration (4th-Order Runge-Kutta, Variant II ODE) ---
    Compute RK4 slopes (k1, k2, k3, k4) using mSEMP_variant2_ode(X_k, A_curr, D_curr)
    Advance state: X(k+1) = X_k + (dt / 6) * (k1 + 2*k2 + 2*k3 + k4)
end for
return X, D_history
```

This attacker -dominant scaling substantially accelerates the infection propagation rate ($\lambda_i$) and immunity loss rate ($\delta_i$) while suppressing the recovery rate ($\gamma_i$). Furthermore, to prevent numerical solver chattering across non-convex equilibrium boundaries and to reflect Security Operation Center (SOC) actuation inertia, Algorithm 3 enforces exponential strategy smoothing with a learning rate of $\alpha = 0.3$, maintaining numerical stability even when $R_0$ remains elevated above the containment threshold.

As illustrated in Fig.6 (Left), inverting the strategic scaling parameter ($\mu > \eta$) prevents complete epidemic eradication and drives the network into a persistent endemic equilibrium over the 150-epoch horizon. Rather than decaying to zero, the infection population I(t) rises steadily to stabilize near 100 nodes (10% of total network capacity), while continuous re-susceptibility ($\delta_i$) caps the recovered population R(t) around 490 nodes. Correspondingly, Fig.6 (Right) demonstrates the defender's adaptive resource reallocation under sustained adversarial stress: budget allocation is progressively shifted from the Gateway tier (declining from 0.23 to 0.12) towards the higher-density Peripheral (0.45-0.50) and Routing (0.40-0.47) tiers to contain the lateral spread. Although minor boundary oscillations emerge between t =40 and t=130 as the solver operates near budget saturation ($K_D = 1.0$), the damped Stackelberg controller successfully averts catastrophic network collapse.

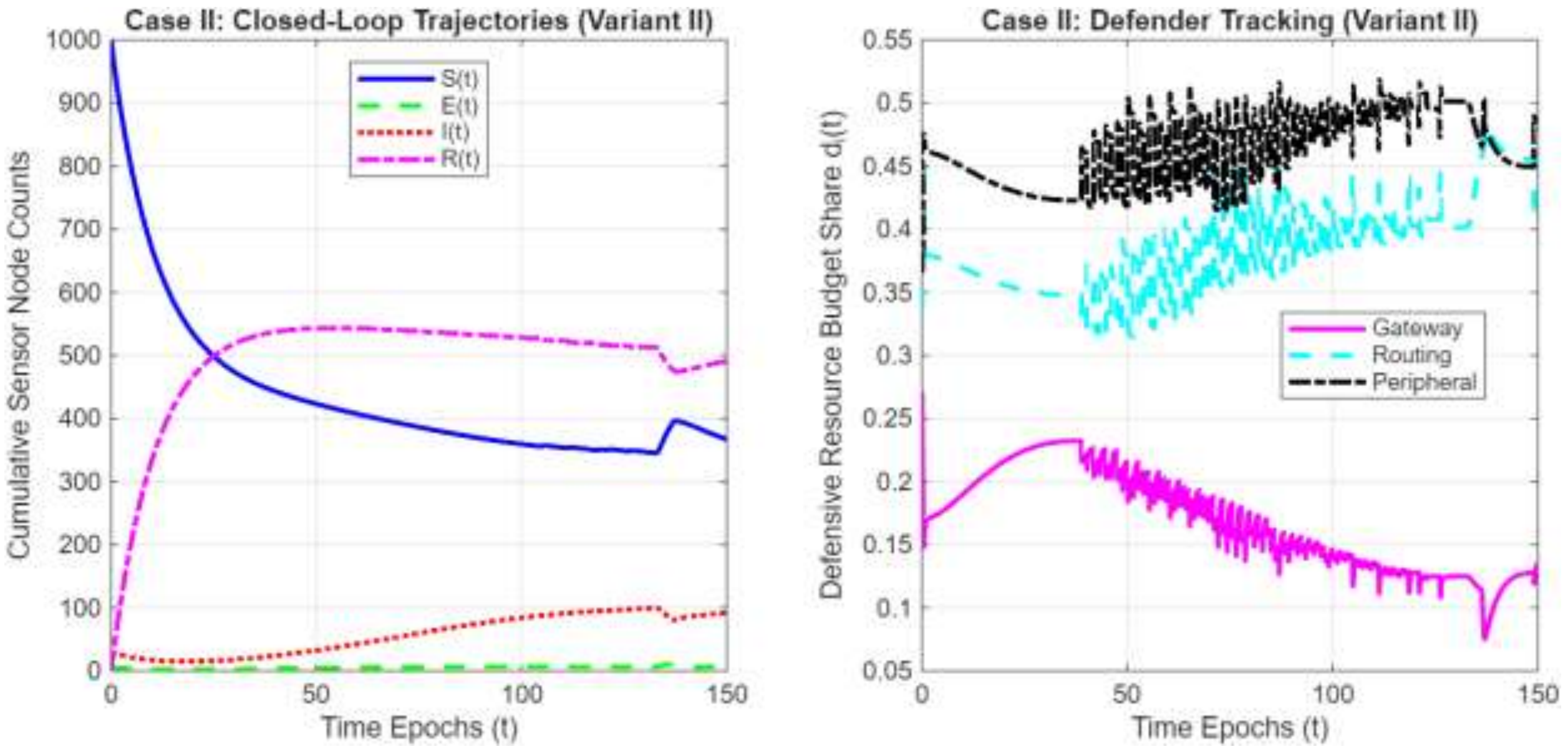

**Fig. 6:** Systemic trajectory under reverted scaling vectors (Variant II). *Left:* Unchecked epidemic cascade suppressed susceptible node due to heightened attacker capabilities. *Right:* Dynamic optimization tracking across heterogeneous node layers under severe network stress.

Table II summarizes the closed loop performance tradeoffs across all four evaluated operational regimes. While Case I establishes a symmetric static baseline (S (50)≈ $455, R(50) \approx 545$) and Case III enforces aggressive perimeter isolation (90% Gateway allocation, restricting compromise via boundary quarantine at the expense of unhardened internal tiers), neither static posture adapts to dynamic adversary pivots. In contrast, Case II (Baseline) dynamically reallocates defensive capital towards the Routing ($\approx 0.47$) and Peripheral ($\approx 0.35$) tiers under KKT stationary, driving active infection I(t) to eradication. Even under severe attacker -dominant scaling ($\mu >$ η) in Case II (Variant II), exponential strategy smoothing ($\alpha = 0.3$), prevents solver chattering and stabilizes the network within a bounded endemic equilibrium (I (150) $\approx 100$) rather than performing catastrophic collapse.

**Table II: Cross- Regime Quantitative Evaluation Summary**

| Evaluation metrics | Case I: Balanced Matchup (Static Baseline) | Case II (Baseline): Dynamic Stackelberg (η ≥ $\mu$) | Case II (Variant II): Attacker Dominant ($\mu$ > η) | Case III: Aggressive Quarantine (Hardened defense) |
|---|---|---|---|---|
| Strategic regime | Symmetric allocation ($A_1 = D_1$) | Defender resilient ($R_o < 1$) | Adversary favored ($\mu_I = 5.0, \mu_R = 6.0$) | Asymmetric perimeter hardening ($D_3$) |
| Control Mechanism | Static proportional allocation | Bi-level SQP Stackelberg | Damped Bi-level SQP ($\alpha$=0.3) | Static Gateway Quarantine |
| Simulation horizon | 50 epochs | 50 epochs | 150 epochs | 50 epochs |
| Terminal Susceptible ($S(T_{final})$) | $\approx$ 455 nodes | $\approx$ 410 nodes | $\approx$ 370 nodes | $\approx$ 355 nodes |
| Terminal Recovered R($T_{final}$)) | $\approx$ 545 nodes | $\approx 555 - 585$ nodes | $\approx$ 490 nodes | $\approx$ 530 nodes |
| Steady-state Infection (I($T_{final}$)) | Contained $I(t) \to 0$ | Eradicated $I(t) \to 0$ | Endemic Equilibrium (I(t) $\approx$ 100) | Contained $I(t) \to 0$ |
| Dominant Tier Allocation | Gateway (0.60) and Routing (0.40) | Routing $\approx$ 0.47 and Peripheral$\approx$ 0.35 | Routing $\approx 0.40 - 0.47$ and Peripheral$\approx 0.45 - 0.50$ | Gateway concentrated (0.90) |
| Budget and Solver Stability | Static | Smooth KKT Convergence | Bounded Saturation Oscillation | Static |

### *VIII. Computational Complexity, Limitations and Real-time Feasibility*

Since the decision variables $A, D \in R^{3x2}$ span a compact six-dimensional simplex across three network tiers and two attack vectors, the bi-level Sequential Quadratic Programming (SQP) formulation exhibits low per epoch dimensionality. Furthermore, starting the solver with damped prior-epoch strategy ($\alpha = 0.3$) restricts local search radius and suppresses oscillatory line search evaluations near non-convex equilibrium boundaries keeping per epoch execution times in the order of milliseconds alongside the four stage RK4 state updates. Since Security Operations Center (SOC) telemetry aggregation and policy enforcement cycles typically operate on second-to-minute timescales, this low computational overhead confirms that GTI-mSEMP framework is well suited for near real-time automated resource orchestration without saturating edge or gateway controller budgets.

While the GTI-MSEMP framework captures closed loop attacker-defender dynamics across heterogeneous tiers, two practical modeling assumptions warrant consideration for operational deployment. First, the bi-level Stackelberg formulation assumes full real-time observability of compartmental node states ($S_i, E_i, I_i, R_i$) via centralized Security and Event Management (SIEM) or Intrusion Detection System (IDS) telemetry; in practice stealthy or polymorphic zero-day payloads may introduce detection latency and partial state observability.

Second defensive resource allocation ($d_{ij}$) are modeled as continuously divisible shares across tiers, whereas physical field deployments often involve discrete actuation constraints such as binary firewall rule toggles or batch firmware updates [30,31]. Addressing partial state observability through stochastic filtering and incorporating integer-constrained policy actuation remain key avenues for future hardware- in- the- loop validation.

## *IX. Conclusion*

This paper presented the Game-*Theory-Integrated Modified Multi- Wireless Sensor Epidemic Malware Propagation (GTI-mSEMP)* framework to model and mitigate strategic, multi-vector malware dissemination across heterogeneous resource constrained cyber-physical networks. By coupling a strictly partitioned three-tier architectural hierarchy (Gateway, Routing and Peripheral layers) with a closed loop bi-level Stackelberg differential game, the framework captures how dynamic attacker-defender resource allocations directly modulate compartmental transition rates and the effective reproduction number ($R_o$). Analytical stability bounds were established via the Next Generation Matrix and Routh-Hurwitz criteria, while the existence and uniqueness of the Subgame-Perfect Stackelberg Equilibrium were proven over compact strategy spaces.

Numerical simulations executed via 4$^{th}$ order Runge-Kutta integration and sequential Quadratic Programming demonstrated that adaptive Stackelberg resource reallocation outperforms static perimeter hardening, driving active infections towards eradication and achieving an approximate 99% recovery rate under balanced operational scaling ($\eta \geq \mu$). Furthermore, sensitivity analysis under an attacker-dominant regime ($\mu > \eta$, Variant II) confirmed that incorporating exponential strategy smoothing ($\alpha = 0.3$) suppresses numerical solver chattering and prevents catastrophic network collapse by progressively shifting defensive capital towards high density Peripheral and Routing tiers stabilizing the network within a bounded endemic equilibrium.

Future work will extend the GTI-mSEMP architecture to incorporate partial state observability via stochastic filtering and discrete integer-constrained policy actuation for hardware-in-the-loop wireless sensor network testbed validation.

**Conflict of Interest**

The authors have no conflict of interest.

**Funding Declaration**

No funding was received for this study**.**